\documentstyle[11pt]{article}
\title{Incompatibility of different customary kaon phase conventions
\thanks{submitted to \it{Phys. Rev. D}}}
\author{Wolfgang J. Mantke\\
{\it CERN, Geneva, Switzerland}\\
and\\
{\it Max-Planck-Institut f\"ur Physik,
Werner-Heisenberg-Institut},\\
{\it F\"ohringer Ring 6,
80805 Munich, Germany}
\thanks{present address, Bitnet: MANTKE@DMUMPIWH.BITNET}}
\begin{document}
\maketitle
\begin{abstract}
The conventions that Wu and Yang assumed
 for the kaon phases in the context of $CP$
symmetrical two-pion decay channels fix the relative kaon phase.
This fact, apparently not
emphasized sufficiently in the past, has recently been overlooked by
Hayakawa and Sanda.  In particular, Wu and Yang fix the relative
phase to a different value than the one resulting from
the convention $CP|K^{0}\rangle = |\overline{K^{0}}\rangle$.
The difference between the two values is made up of possible
contributions from $CPT$- and direct $CP$-violations
during the decay of a kaon into a two-pion state of isospin zero.
\end{abstract}

I would like to comment on the phase freedom in the neutral kaon system
as presented in a recent paper by Hayakawa and
Sanda \cite{ci:HayakawaSanda}.
Following Lee and Wolfenstein \cite{ci:LeeWolfenstein}
I assume that the $CP$ operator is fixed.  An argument similar to theirs
applies to the $T$ operator.

In order to keep different kaon phase conventions apart, one should be
aware of a convention that is hardly ever mentioned
explicitly but nonetheless employed by most authors.
For two-pion states, the isospin phase convention is assumed which is
implicit in the Clebsch-Gordon series.   Because the Clebsch-Gordon
coefficients are real, the isospin vectors
$|(2\pi)_{I}\rangle$ ($I=0,2$ \& $I_{3}=0$) as well as their
combinations $|\pi^{+}\pi^{-}\rangle$ and
$|\pi^{0}\pi^{0}\rangle$ acquire the same
phase factor under the $CPT$ transformation.  The kaon phases are chosen
such that the transformation of the kaon states is associated with
this phase factor, too,
\begin{equation}
\label{eq:CPTonkzeroandpion}
CPT|K^{0}\rangle=e^{i\theta}|\overline{K^{0}}\rangle,\qquad\qquad
CPT|(2\pi)_{I}\rangle=e^{i\theta}|(2\pi)_{I}\rangle.
\end{equation}

Note that because $CPT$ is idempotent on integral spin systems and
antilinear, the transformation of the antikaon state vector is associated
with the same phase factor,
\begin{equation}
CPT|\overline{K^{0}}\rangle=e^{i\theta}|K^{0}\rangle.
\end{equation}
I call the relative phase between the state vectors of the
kaon and the antikaon briefly ``the relative kaon phase''.  This phase
is not fixed by the value of $\theta$.
In the convention of Eq.\ (\ref{eq:CPTonkzeroandpion}),
$CPT$ invariance of the two-pionic decay
amplitudes (without strong scattering phase shift) $A_{I}$
of the kaon and $\overline{A}_{I}$ of the antikaon
takes on the form,
\begin{equation}
\label{eq:CPTinvarianceofA}
CPT \mbox{ invariance}\qquad \Rightarrow\qquad
\overline{A}_{I} = A_{I}^{*}.
\end{equation}
Hayakawa and Sanda, too, stated this as a consequence of $CPT$
invariance \cite{ci:HayakawaSanda1}
and therefore surely had assumed the same phase condition.

Wu and Yang \cite{ci:WuYang} employed two conventions:
First they assumed the kaonic and pionic $CPT$ phase factors to be
equal, as done in Eq.\ (\ref{eq:CPTonkzeroandpion});
second they chose $A_{0}$ to be positive,
\begin{equation}
\label{eq:WuYangconvention}
 A_{0}^{WY} =\mbox{real}> 0,
\end{equation}
which fixes the relative phase between the $K^{0}$ and the $(2\pi)_{0}$
state vectors. Eq.s (\ref{eq:CPTonkzeroandpion}) and
(\ref{eq:WuYangconvention}) {\it together} fix the relative kaon
phase.

Schubert et al. \cite{ci:Schubert1}
generalized the second convention of Wu and Yang into
\begin{equation}
\label{eq:Schubert}
\frac{A_{0}^{S}}{\overline{A}_{0}^{S}}= \mbox{real} > 0,
\end{equation}
which fixes the relative kaon phase.

{}From $CPT|K^{0}\rangle=TCP|K^{0}\rangle$ follows that the phase
factors under the $CP$ transformation and time reversal are
related as follows \cite{ci:Nakada},
\begin{eqnarray}
CP|K^{0}\rangle&=&e^{i\phi}|\overline{K^{0}}\rangle,\qquad\quad
CP|\overline{K^{0}}\rangle=e^{-i\phi}|K^{0}\rangle,\nonumber\\
T|K^{0}\rangle&=&e^{i\chi}|K^{0}\rangle,\qquad\qquad
T|\overline{K^{0}}\rangle =
e^{i\overline{\chi}}|\overline{K^{0}}\rangle,\nonumber
\end{eqnarray}
\begin{equation}
\label{eq:CP&Tandphase}
2\phi=\overline{\chi}-\chi.
\end{equation}

In order to analyze the conventions of Wu and Yang, and of Schubert,
I shall now derive a formula (Eq.\ (\ref{eq:formulaforphi})\,)
for the $CP$ phase factor $e^{i\phi}$.

The mixing eigenvectors can be expanded as
\begin{equation}
\label{eq:alphaKSKL}
|K_{S,L}\rangle
=\frac{e^{i\alpha_{S,L}}}{\sqrt{2(1+|\tilde{\varepsilon}\pm\delta|^{2})}}
[(1+\tilde{\varepsilon}\pm\delta)|K^{0}\rangle\pm
(1-\tilde{\varepsilon}\mp\delta)|\overline{K^{0}}\rangle ].
\end{equation}
I choose the relative phases between the state vector of the $K^{0}$,
the $CP$ eigenvectors $|K_{1,2}\rangle$
and the mixing eigenvectors such that
\begin{eqnarray}
\label{eq:standardK1K2}
|K_{1,2}\rangle &=& \frac{1\pm CP}{\sqrt{2}}|K^{0}\rangle,\\
\label{eq:standardKSKL}
|K_{S,L}\rangle
&=& \frac{1}{\sqrt{1+|\varepsilon_{S,L} |^{2}}}
(|K_{1,2}\rangle + \varepsilon_{S,L}  |K_{2,1}\rangle ).
\end{eqnarray}
The only remaining phase freedom of the state vectors
lies in the relative kaon phase, which we represent by the $CP$
phase $\phi$ of Eq.\ (\ref{eq:CP&Tandphase}).
Only if we choose the relative kaon phase according to
\begin{equation}
\label{eq:LeeOehmeYang}
e^{i\phi^{CP}}=1,
\end{equation}
can we obtain the relations
\begin{equation}
\label{eq:epsilonanddelta}
\varepsilon_{S,L} =\tilde{\varepsilon}^{CP}\pm\delta^{CP},
\qquad\qquad e^{i\alpha_{S,L}^{CP}}=1.
\end{equation}
I shall refer to the phase choice in Eq.\ (\ref{eq:LeeOehmeYang}) as
the ``charge-parity convention''.
Under a change of the relative kaon phase, the ratio of the coefficient
in front of $|K^{0}\rangle$ with the one in front of
$|\overline{K^{0}}\rangle$ transforms inversely to the $CP$ phase factor
$e^{i\phi}$ of Eq.\ (\ref{eq:CP&Tandphase}).  Therefore the product of
this ratio and this phase factor is invariant, so that
the new, primed quantities are related to the old ones by
\begin{equation}
\frac{1+(\tilde{\varepsilon}\pm\delta)}
{1-(\tilde{\varepsilon}\pm\delta)}e^{i\phi}
=\frac{1+(\tilde{\varepsilon}'\pm\delta')}
{1-(\tilde{\varepsilon}'\pm\delta')}e^{i\phi'},
\qquad\Rightarrow
\end{equation}
\begin{eqnarray}
\tilde{\varepsilon}'\pm\delta'&=&\frac
{(\tilde{\varepsilon}\pm\delta)\cos\frac{\phi'-\phi}{2}
-i\sin\frac{\phi'-\phi}{2}}{\cos\frac{\phi'-\phi}{2}
- i(\tilde{\varepsilon}\pm\delta)\sin\frac{\phi'-\phi}{2}}\nonumber\\
&=&(\tilde{\varepsilon}\pm\delta)-i\frac{\phi'-\phi}{2}
+O([\tilde{\varepsilon}\pm\delta][\phi'-\phi],
[\phi'-\phi]^{2}).
\end{eqnarray}
Hence $\delta$ and $\Re\tilde{\varepsilon}$ are phase invariant to the
first order, whereas $\Im\tilde{\varepsilon}$ is phase dependent in all
orders.  Furthermore
\begin{equation}
\label{eq:coefftransf}
1+\tilde{\varepsilon}'\pm\delta'=e^{-i\frac{\phi'-\phi}{2}}
\frac{1+\tilde{\varepsilon}\pm\delta}{\cos\frac{\phi'-\phi}{2}
-i(\tilde{\varepsilon}\pm\delta)\sin\frac{\phi'-\phi}{2}}.
\end{equation}
Because the relative phases between $|K^{0}\rangle$ and
$|K_{S,L}\rangle$ have been fixed by Eq.s\
(\ref{eq:standardK1K2}) and (\ref{eq:standardKSKL}),
the phase factors in Eq.\ (\ref{eq:alphaKSKL}) transform inversely to
the coefficients in front of  $|K^{0}\rangle$,
\begin{equation}
e^{i\alpha_{S,L}'}= e^{i\frac{\phi'-\phi}{2}}e^{i\alpha_{S,L}} +
O([\tilde{\varepsilon}\pm\delta][\phi'-\phi],[\phi'-\phi]^{2}).
\end{equation}
With this equation we can pass from the convention of Eq.s
(\ref{eq:LeeOehmeYang}) and (\ref{eq:epsilonanddelta})
to a convention close to it ($\phi$ small).  We find that
to the first order the phase factors of the two mixing eigenstates are
equal to one another and given by
\begin{equation}
\label{eq:alphaphase}
e^{i\alpha_{S,L}}= e^{i\frac{\phi}{2}}
+ O(\varepsilon_{S,L} \phi,\phi^{2}).
\end{equation}

The ratios \cite{ci:Barmin}
\begin{equation}
\label{eq:epnew&a}
\varepsilon:=\frac{\langle (2\pi)_{0}|{\cal T}|K_{L}\rangle}
{\langle (2\pi)_{0}|{\cal T}|K_{S}\rangle},\qquad\qquad
a:=\frac{\langle (2\pi)_{0}|{\cal T}|K_{2}\rangle}
{\langle (2\pi)_{0}|{\cal T}|K_{1}\rangle}
\end{equation}
are related by
\begin{equation}
\label{eq:epnew&aII}
\varepsilon=\frac{a+\varepsilon_{L}}{1+\varepsilon_{S} a},
\qquad\Rightarrow\qquad
a=\frac{\varepsilon-\varepsilon_{L}}{1-\varepsilon\,\varepsilon_{S}}.
\end{equation}
Most authors, e.g.\ \cite{ci:HayakawaSanda}
\cite{ci:Schubert1}
\cite{ci:MaSimmonsTuan}
 \cite{ci:Commins},
use different relative phases between $|K_{S}\rangle$,
$|K_{L}\rangle$ and
$|K^{0}\rangle$ in that they let the phase factors of Eq.\
(\ref{eq:alphaKSKL}) be equal to unity.  (The inverses of my phase
factors then would need to appear on the right side of Eq.\
(\ref{eq:standardKSKL}).)\,  My parameter
$\varepsilon$ is equal to their
$\langle (2\pi)_{0}|{\cal T}|K_{L}\rangle$/
$\langle (2\pi)_{0}|{\cal T}|K_{S}\rangle$
 times the factor
$e^{i(\alpha_{L}-\alpha_{S})}$, which is unity to the first order.
$a$ is a parameter of direct $CP$ violation in the kaon-decay channel
with final state $(2\pi)_{0}$.  The expression \cite{ci:Schubert1}
\begin{equation}
\label{eq:alphazero}
\alpha_{0} := \frac{A_{0}-\overline{A}_{0}}{A_{0}+\overline{A}_{0}}
\end{equation}
becomes a $CPT$ violation parameter of the same channel, when one
uses the conventions of Wu and Yang or the one of Schubert et al..

For the $CP$ phase as fixed by Wu and Yang, or by Schubert
we now find the formulas
\begin{eqnarray}
\label{eq:formulaforphi}
e^{i\phi^{WY,S}} &=& \frac{\overline{A}_{0}^{CP}}{A_{0}^{CP}}
\,\frac{A_{0}^{WY,S}}{\overline{A}_{0}^{WY,S}}
=\frac{1-a}{1+a}\,\frac{1+\alpha_{0}^{WY,S}}{1-\alpha_{0}^{WY,S}},\\
\Rightarrow\quad \phi^{WY,S} &=& 2 \Im (-a+\alpha_{0}^{WY,S})
+ O(a^{2}, [\alpha_{0}^{WY,S}]^{2}, a\alpha_{0}^{WY,S}),
\end{eqnarray}
where the superscript $CP$ refers to the charge-parity convention,
 Eq.\ (\ref{eq:LeeOehmeYang}).
$CPT$ symmetry would imply,
\begin{equation}
CPT\mbox{ invariance}\quad\Rightarrow\quad
e^{i\phi^{WY}}=e^{i\phi^{S}}=e^{-i2\arg (A_{0}^{CP})}
=\frac{1-a}{1+a},
\end{equation}
where $A_{0}^{CP}$ is the value of $A_{0}$, if conventions
Eq.\ (\ref{eq:CPTonkzeroandpion})  and Eq.\ (\ref{eq:LeeOehmeYang})
are jointly assumed.
(We may supplement the convention of Schubert, Eq.\ (\ref{eq:Schubert}),
with the one of Eq.\ (\ref{eq:CPTonkzeroandpion}).
In the case of $CPT$ invariance $A_{0}$
then would be real.  If $A_{0}$ were
positive we would have arrived at the conventions of Wu and Yang;
otherwise the relative phases between the state vectors of the two-pion
system and the ones of the kaon system would be opposite to
the corresponding choices of Wu and Yang.)

Assuming $CPT$ symmetry and both conventions of Wu and Yang,
$\tilde{\varepsilon}$
is equal to the $CP$ violation parameter $\varepsilon$
\cite{ci:Schubert2},
\begin{equation}
CPT\mbox{ invariance}\quad\Rightarrow\quad
\tilde{\varepsilon}^{WY}
=\varepsilon e^{i(\alpha_{S}^{WY}-\alpha_{L}^{WY})}
=\varepsilon [1
+O(\varepsilon^{2},\varepsilon_{S,L}\varepsilon_{S,L},\varepsilon\,
\varepsilon_{S,L})  ].
\end{equation}

The first order expression for $\tilde{\varepsilon}$ in terms of the
off-diagonal elements of the kaon matrix ${\cal M}$ and its eigenvalues
$\lambda_{S,L}$ is
\begin{equation}
\tilde{\varepsilon}_{\cal M} := \frac{
\langle\overline{K^{0}}|{\cal M}|K^{0}\rangle-
\langle
K^{0}|{\cal M}|\overline{K^{0}}\rangle}{2(\lambda_{L}-\lambda_{S})}
= \tilde{\varepsilon} +
O(\tilde{\varepsilon}^{2},\delta^{2},\tilde{\varepsilon}\delta);
\end{equation}
{\it only} if $e^{i2\phi}=1$ (and hence
$e^{i\chi}=e^{i\overline{\chi}}$, see Eq.\ (\ref{eq:CP&Tandphase}))
does $\tilde{\varepsilon}_{\cal M}$ measure $CP$ and $T$ violation
of the kaon matrix \cite{ci:MaSimmonsTuan}.

At  the very end of their appendix C, Hayakawa and
Sanda \cite{ci:HayakawaSanda} assume the convention
$CP|K^{0}\rangle = |\overline{K^{0}}\rangle$ (i.e. $\phi=0$) and the ones
of Wu and Yang simultaneously.  This slip, which had already been
made elsewhere, e.g.\ \cite{ci:Commins} \cite{ci:Sachs},
apparently did not propagate to anywhere else in their paper.

A related confusion arose when several authors \cite{ci:Barmin}
 \cite{ci:Cronin} \cite{ci:Tanner}
did not point out that
$\tilde{\varepsilon}$ (or $\tilde{\varepsilon}_{\cal M}$) ceased to be a
$T$ violation parameter of the kaon matrix, when they switched from the
charge-parity convention, $\phi^{CP}=0$, to the one of Schubert,
Eq. (\ref{eq:Schubert}).
Moreover, it was overlooked \cite{ci:Barmin1}
that these two conventions are incompatible.

\vspace{6 mm}

I am grateful to the members of the CPLEAR collaboration for their
hospitality and fruitful discussions, especially with
Maria Fidecaro, Tatsuya Nakada and Bernd Pagels.  Moreover I thank
Giancarlo D'Ambrosio for commenting on the manuscript.
%
%========================================================================
\newpage


\begin{thebibliography}{99}
\bibitem{ci:HayakawaSanda2}
M. Hayakawa and A.I. Sanda, ref. \cite{ci:HayakawaSanda},
appendix  C, 1163.\\
Note that my Eq.\ (\ref{eq:CPTinvarianceofA}), and
Eq.\ (C7) at the end of this appendix only hold, if
the first convention of Wu and Yang
(my Eq.\ (\ref{eq:CPTonkzeroandpion})) is assumed.
\end{thebibliography}
\end{document}